\documentclass[preprint,12pt]{elsarticle} 

\usepackage{amssymb}

\usepackage{graphicx}
\usepackage{xcolor}

\newcommand{\rf}[1]{(\ref{#1})}

\newcommand{\beq}{\begin{equation}}
\newcommand{\eeq}{\end{equation}}
\newcommand{\beqr}{\begin{eqnarray}}
\newcommand{\eeqr}{\end{eqnarray}}
\newcommand{\lb}[1]{\label{#1}}
\newcommand{\bc}{\begin{center}}
\newcommand{\ec}{\end{center}}
\newcommand{\ct}[1]{\cite{#1}}

\journal{Photonics and nanostructures ... }

\begin{document}

\begin{frontmatter}



\title{Spontaneous emission, collective phenomena and the efficiency of plasmon-stimulated photoexcitation.}


\author[inst1]{Igor E. Protsenko}

\affiliation[inst1]{organization={P.N.Lebedev Physical Institute of the RAS},
            addressline={Leninsky prospect 53}, 
            city={Moscow},
            postcode={119991}, 
            country={Russia}}

\author[inst1]{Alexander V. Uskov}
\author[inst2]{Nikolay V.~Nikonorov}

\affiliation[inst2]{organization={ITMO University},
            addressline={Kronverksky Pr. 49}, 
            city={St.~Petersburg},
            postcode={197101}, 
            country={Russia}}

\begin{abstract}
We find that the spontaneous and collective emissions have a strong influence on the excitation of two-level absorbers (atoms, molecules) interacting in resonance with the plasmonic mode near the metal nanoparticle. The spontaneous and collective emissions limit the absorption enhancement by the plasmonic mode and make the enhancement possible only with a fast, picosecond population relaxation of the upper absorbing states. Conditions for the maximum of plasmon-enhanced absorption in the presence of spontaneous and collective emissions are found. The nonlinearity in the nanoparticle-absorber interaction and in collective emission causes the bistability in the plasmon-enhanced absorption at high external field intensities and the plasmonic mode excitation.
\end{abstract}




\end{frontmatter}


%
\section{\label{Sec1}Introduction}
Excitation of localised plasmon resonances (LPR) leads to electric field enhancement in the vicinity of metal nanoparticles  \ct{Maier2007,https://doi.org/10.1002/lpor.200810003,Amendola_2017,doi:10.1021/jp8060009}. Such a "plasmonic" field enhancement finds various applications in optical sensors \ct{C4AN01079E}, bio-medicine  \ct{doi:10.2217/17435889.1.2.201}, photovoltaics \ct{Jang2016}, photocatalysis \ct{Zhang_2013} and  other areas of physics and chemistry \ct{Maier2007,ma12091502}. A new branch of chemistry - plasmonic chemistry - has been developed around the chemical reactions \ct{Zhan2023,https://doi.org/10.1002/adom.201700191} stimulated by the electric field enhancement near plasmonic nanoparticles. 

A photochemical reaction is initiated by the absorption of light and the creation of transient molecules in excited states whose properties differ greatly from the original molecules.  These new chemical species can fall apart, change into new structures, combine other molecules, or transfer electrons, hydrogen atoms, protons, or their electronic excitation energy to other molecules \ct{En_Br}. Therefore, it is important to optimise the plasmonic photochemical reactions by searching for conditions of maximum field and absorption enhancement near metal nanoparticles \ct{doi:10.1021/jp026731y}. This work contributes to such optimization.

A large electric field enhancement is known to occur near the sharp features of nanostructures, known as 'hot spots' \ct{Díaz-Núñez2019}, or in a small gap between nanoparticles \ct{B917543A}. 
A plasmonic gap field mode has been shown to have a high potential for achieving ultra-strong light absorption, the magnitude of which can be relatively easily controlled \ct{Baumberg2019,Yeshchenko2020,Huang2016,Lumdee2014}. In this paper we consider, as a first approximation, a much simpler system, limited to the near-field absorption enhancement of the isolated metal nanoparticle with the excited LPR.  The approach and results of the present simple model will help to analyse the perspective but more complicated problem of absorption enhancement in plasmonic gaps.     
 
We will focus on the spontaneous and collective emissions, which significantly affect the resonant absorption near plasmonic nanoparticles and lead to new physical effects.  Spontaneous emission near metal nanoparticles has been studied in many papers, for example in \ct{BLANCO200437}. Collective effects in the form of superradiance have been studied for the resonant absorbers, such as atoms or molecules, in the vicinity of the plasmonic particle \ct{Shahbazyan2011,Protsenko_2015,Protsenko_2017}.  However, the absorption enhancement in the vicinity of the nanoparticle, taking into account the spontaneous and collective emissions, deserves further study. The influence of spontaneous emission and collective effects on absorption is important for plasmonic chemistry, especially in the context of the recently studied plasmon-induced direct resonance energy transfer (PIRET) processes \ct{10.1063/1.5050209}. Part of the study of plasmon-enhanced absorption with spontaneous and collective emission in the LPR mode, with direct energy transfer from the plasmonic nanoparticle to the absorbing molecule, is carried out in the present work.

Here we use the model for resonant absorption near plasmonic nanoparticles, similar to the laser model. The plasmonic nanoparticle forms a cavity for the electric field similar to the laser cavity as in the plasmonic nanolaser \ct{PhysRevA.71.063812,IEProtsenko_2008,IEProtsenko_2012_UFN}. Part of our approach is similar to the analysis of spontaneous emission and collective effects in lasers \ct{Andre:19,Protsenko_2022}.  

The plasmonic mode provides good conditions for collective effects in the resonant absorption. Collective effects in the resonant medium are significant in the low quality cavity \ct{Andre:19,Protsenko_2022,PhysRevA.105.L011702}. The LPR mode is of low quality. The particle surface is a good location for the resonant absorbers (atoms, molecules, q-dots, etc.). 

In this paper, we show that the spontaneous emission in the LPR mode and the collective interaction of absorbers through the plasmonic electric field significantly affect the absorber excitation in the vicinity of metal nanoparticles and lead to bistability in the excitation. We will find conditions for absorber excitation enhancement and bistability. These results are important for all branches of plasmonic chemistry, photovoltaics and all applications related to resonant excitation of absorbers near plasmonic nanoparticles.

We derive the fundamental equations quantum mechanically as Heisenberg operator equations. In the present work, we solve these equations in the semi-classical approximation, as it is used for example in laser theory \ct{SLW}. A full quantum analysis of the equations will be performed in the future.

We perform analytical calculations. The analytical approach allows us to find expressions that illustrate the physical processes at the excitation of the absorber and the influence of the plasmonic particle on important parameters such as the polarisation and the population of states of the absorber.

We consider the increase in the absorber excited state population as a criterion for absorption enhancement by the LPR. We describe the interaction of the dipole absorbers (molecules) with the dipole mode of the spherical plasmonic nanoparticle. According to Mie theory \ct{Mie}, the external field of the nanoparticle dipole mode is the field of the dipole in the nanoparticle center. Thus, the nanoparticle-absorber interaction is the dipole-dipole interaction as long as the high-order plasmonic mode excitation is neglected. We assume the resonance between the absorber transition and the frequency of the plasmonic dipole mode.  The resonant dipole-dipole interaction at a short absorber-nanoparticle distance  causes the well-known Förster nonradiative resonance energy transfer (FRET)  \ct{Helms2008}. Thus, in our work, the effect of collective and single absorber spontaneous emission in the plasmonic mode on the FRET process is highlighted.  

We assume that the molecules are located on the nanoparticle dielectric shell with a thickness of at least 5 nm, which is sufficient to avoid strong absorber fluorescence quenching  \ct{PhysRevLett.96.113002}.  

We assume that the nanoparticle radius is $13$~nm, so the total distance of $18$~nm between the nanoparticle center and the absorbers on the shell is much smaller than the optical wavelength of the resonant field $\sim 1~\mu$m, see the scheme in Fig.~\ref{Fig1}. We therefore use the quasi-static approximation for the absorber-nanoparticle dipole-dipole interaction.

We take into account the nonradiative and radiative \ct{Protsenko_2012} losses of the nanoparticle, including the losses associated with the collisions of the metal electrons with the nanoparticle surface \ct{Protsenko_2012, Uskov2014}. Thus, such a model based on the dipole-dipole interaction of the absorbers with the nanoparticle is a good approximation for our purposes. We note that more sophisticated models, such as macroscopic QED for quantum nanophotonics \ct{Garcia-Vidal_2021} and other approaches as discussed in the review \ct{Nguyen_2015}, can be used for advanced investigation of the spontaneous and collective emission in plasmonic nanostructures outlined in this paper. A relatively simple model given below is  useful for finding similarities in the collective emission phenomena in small lasers, LEDs and resonant absorber plasmonic structures.

In the next section, we describe the model, the Hamiltonian, and the equations of motion.

The expressions characterising the steady-state absorption, such as the absorption coefficients and the energy balance equation, are presented in section~\ref{sec3}. Several physical processes in the resonant absorption near the metal nanoparticle are explained with the help of the equations found in section~\ref{sec3}.

Section~\ref{sec4} shows the conditions for the best plasmon-enhanced absorption and describes the bistability.

Examples of the absorption enhancement and the bistability are shown in section~\ref{sec5} for typical values of the nanoparticle and the absorber parameters.

The results are discussed in section~\ref{sec6} and the paper is concluded in section~\ref{sec7}.  
\section{ The model, Hamiltonian and equations of motion}\label{section2}
\subsection{The model}
\begin{figure}[thb]\bc
\centering
\includegraphics[width=9cm]{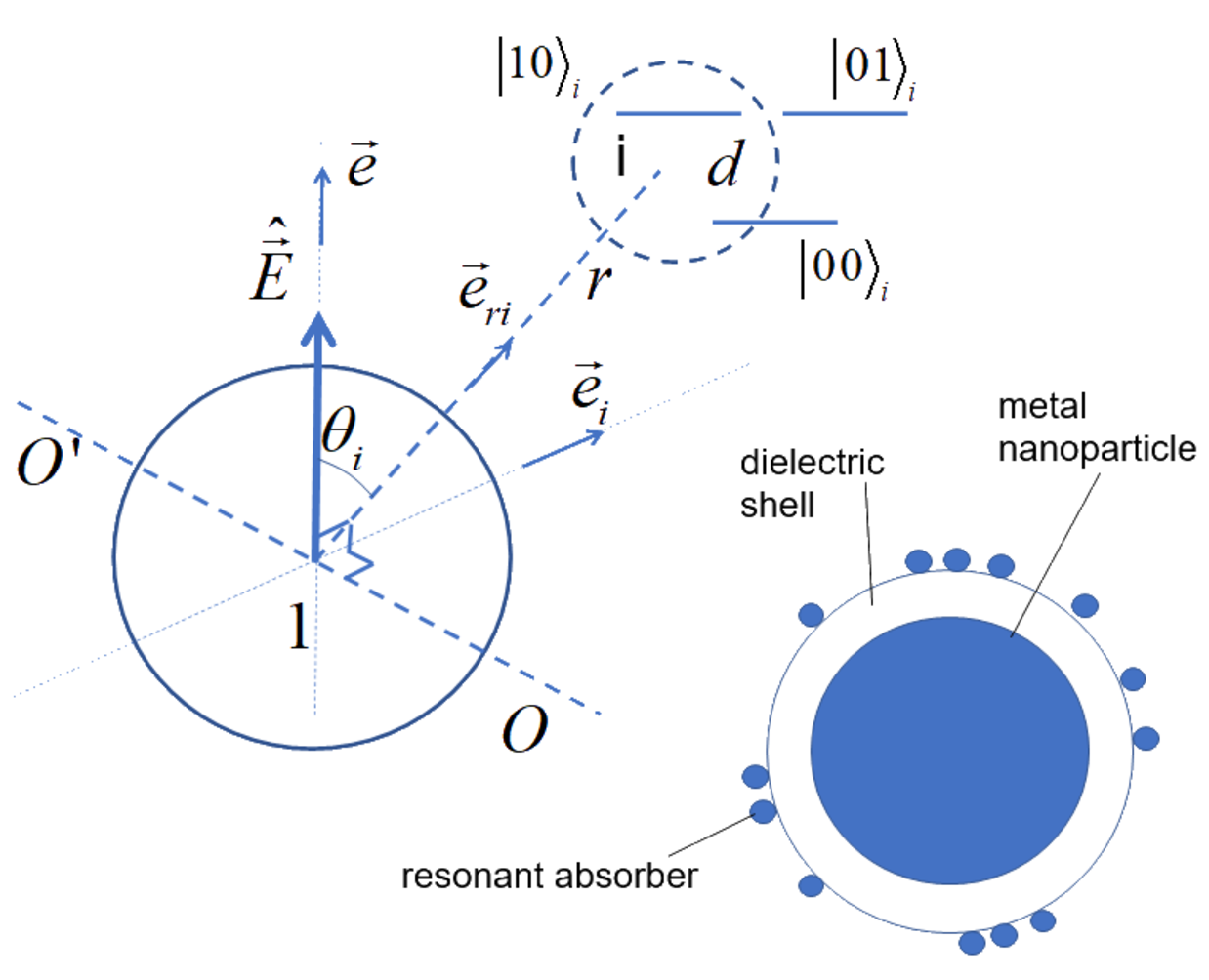}
\caption{Coordinate system, the nanoparticle 1 and the resonant absorber $i$ (a molecule or an atom) in the external electric field $\hat{\vec{E}}$. $r$ is the distance, and $\vec{e}_{ri}$ is the unit vector from the centre of the nanoparticle   to the absorber. $\left|00\right>_i$ is the ground state of the absorber. The dipole momentum with the matrix element $d$ for the absorber transition from the excited state $\left|01\right>_i$ (or $\left|10\right>_i$) to $\left|00\right>_i$ is parallel to the unit vector  $\vec{e}$ of the external field polarisation (or to the vector $\vec{e}_i$). The absorber state with the transition dipole momentum parallel to the axis $OO'$ is not excited; such transition does not interact with the external and the nanoparticle fields. The inset shows the resonant absorbers randomly distributed on the nanoparticle shell.  }
\label{Fig1}\ec
\end{figure}

Suppose, a metal nanoparticle interacts with $i=1...N_m$ identical absorbers, as molecules or atoms, located at the distance $r$ from the particle center on the dielectric shell as shown in Fig.~\ref{Fig1}. The transition frequency of the absorber is in resonance with the dipole LPR mode of the nanoparticle. 
%
%
%
%
Such resonance can be achieved for the nanoparticle of appropriate shape \ct{doi:10.1021/jp026731y}. For simplicity, we will consider a spherical nanoparticle. A metallic nanoparticle with resonant absorbers/emitters on the dielectric shell is a well-known object in plasmonic theory and experiments.
It has been considered and realised, for example, in the plasmonic nano-lasers \ct{Stamplecoskie2014,Noginov2009,IEProtsenko_2012_UFN}, for the studies of the plasmonic superradiance \ct{PhysRevB.82.075429,Protsenko_2015} and, most recently, in the discovery of the role of electron spillout in strong coupling of the plasmonic nanoparticle with the excitonic shell \ct{bundgaard2023quantuminformed}.

The particle and the absorbers interact with the external monochromatic electric field of the amplitude $\hat{\vec{E}}$  linearly polarised along the unit vector $\vec{e}$. The field $\hat{\vec{E}}$ is in resonance with the LPR and the absorber transitions. 

We consider the $i$-th absorber in the Cartesian coordinate system with the origin at the particle centre. One coordinate axis is parallel to $\vec{e}$, another axis with the unit vector $\vec{e}_i$ is in the plane formed by $\vec{e}$ and the unit vector $\vec{e}_{ri}$ points from the nanoparticle centre to the i-th absorber.  The angle $\theta_i$ is between $\vec{e}$ and $\vec{e}_i$, and the third  axis of coordinate is along the line $OO'$. Two excited ${{\left\langle  10 \right|}_{i}}$, ${{\left\langle  01 \right|}_{i}}$ and the ground ${{\left\langle  00 \right|}_{i}}$ states of the i-th absorber participate  interact with electric fields near the nanoparticle. Such absorber states are, for example, the lowest unoccupied molecular orbital (LUMO) and the highest occupied molecular orbital (HOMO)   \ct{10.1063/1.1700523}.

We will calculate the population of excited states of the absorbers and study how the interaction between the particle and the absorbers affects the population. 

We assume a large number $N_m\gg 1$ of absorbers randomly distributed on the nanoparticle shell, as shown in the inset in Fig.~\ref{Fig1}. In the case of $N_m\gg 1$, the absorber-absorber interaction through the near-field leads only to the self-broadening  of the absorber's transitions \ct{doi:10.1139/p75-298}. We include the self-broadening in the absorber transition width $\gamma$ and do not explicitly describe the absorber-absorber interaction. 
\subsection{Hamiltonian}
Following \ct{PhysRevA.71.063812}, we describe the nanoparticle as a dipole that undergoes harmonic oscillations in the resonant external electric field. We introduce the nanoparticle dipole momentum operator 
\beq
\hat{\vec{p}}=p\vec{e}\left( \hat{b}{{e}^{-i{{\omega }_{0}}t}}+{{{\hat{b}}}^{+}}{{e}^{i{{\omega }_{0}}t}} \right) \lb{dip_mom_op}
\eeq
where $p$ is the dipole momentum matrix element, $\hat{b}$ is a Bose-operator describing collective (plasmonic) oscillations of the nanoparticle electrons in the electric field. Later we express $p$ in terms of the polarisability of the nanoparticles \ct{PhysRevA.71.063812}. For reasons of symmetry, we assume the same directions for the nanoparticle dipole and the external field polarisation.

The dipole momentum operator of the $i$th absorber resonance transition is
\beq
{{\hat{\vec{d}}}_{i}}=d\left[ \left( {{{\hat{\sigma }}}_{i}}\vec{e}+{{{\hat{\sigma }}}_{\bot i}}{{{\vec{e}}}_{i}} \right){{e}^{-i{{\omega }_{0}}t}}+h.c. \right],\lb{dip_mom_mol_i}
\eeq
where $d$ is the dipole transition matrix element; absorber operators in Dirac notation   ${{\hat{\sigma }}_{i}}={{\left| 00 \right\rangle }_{i}}{{\left\langle  01 \right|}_{i}}$,  (or ${{{\hat{\sigma }}}_{\bot i}}={{\left| 00 \right\rangle }_{i}}{{\left\langle  10 \right|}_{i}}$) are transition operators from the high $\left|01\right>_i$ (or $\left|10\right>_i$) to the low $\left|00\right>_i$ states of the absorber $i$ parallel to $\vec{e}$, (or $\vec{e}_i$): ${{{\hat{\sigma }}}_{i}}\left|01\right>_i=\left|00\right>_i$ and ${{{\hat{\sigma }}}_{\perp i}}\left|10\right>_i=\left|00\right>_i$.

The Hamiltonian of the particle and $N_m$ absorbers in the electric field $\hat{\vec{E}}$ is 
\beq
H={{H}_{0}}-\left( \hat{\vec{p}}+\sum\limits_{i=1}^{N}{{{{\hat{\vec{d}}}}_{i}}} \right)\hat{\vec{E}}+{{\hat{V}}_{int}}+\hat{\Gamma }\lb{Ham_0}
\eeq
The first term in \rf{Ham_0} describes the free oscillations of the dipoles; the second term is for the dipole interaction of the particle and absorbers with the external field; the third term ${{\hat{V}}_{int}}$ describes the interaction of the particle and absorbers through the nanoparticle near field (the LPR field), and the last term in \rf{Ham_0} describes the dissipation and the interaction of the particle and absorbers with baths. 

We assume that the distance $r$ between the absorbers and the particle centre is much smaller than the LPR wavelength. Thus the particle-absorber interaction  ${{\hat{V}}_{int}}$  is a quasi-static dipole-dipole interaction. We take the quasi-static dipole-dipole interaction energy  from \ct{Landau_FT}, replace the classical dipoles by operators, and write
\beq
{{\hat{V}}_{int}}=\frac{1}{{{r}^{3}}}\sum\limits_{i=1}^{{{N}_{m}}}{\left[ \hat{\vec{p}}{{{\hat{\vec{d}}}}_{i}}-3(\hat{\vec{p}}{{{\vec{e}}}_{ri}})({{{\hat{\vec{d}}}}_{i}}{{{\vec{e}}}_{ri}}) \right]},\lb{dd_int}
\eeq
where ${{{\vec{e}}}_{ri}}$ is the unit vector along the direction from the nanoparticle centre to the ith absorber, as shown in Fig.~\ref{Fig1}.

The external field is
\beq
\hat{\vec{E}}={{E}_{0}}\vec{e}\left( \hat{a}{{e}^{-i{{\omega }_{0}t}}}+{{{\hat{a}}}^{+}}{{e}^{i{{\omega }_{0}}t}} \right) \lb{ext_f}
\eeq
where $\hat{a}$ is the field amplitude Bose operator, ${{E}_{0}}=\eta^{-1}\sqrt{{2\pi \hbar {{\omega }_{0}}}/{V}}$,  $V$ is the field quantization volume ($V$  is not used in the final results \ct{Lifshitz1982a}), $\eta$ is the refractive index of the nanoparticle environment \ct{Tkalya2000}.  We assume that the distances between the absorbers and the particle centre are much smaller than the wavelength of the external field, so that the phase (assumed to be zero) of the external field is the same for all absorbers and the particle.

Substituting the  expressions \rf{dip_mom_op}, \rf{dip_mom_mol_i} into the sum in Eq~\rf{dd_int}, dropping the fast oscillating terms $\sim e^{\pm 2i\omega_0t}$, we get
\beq
\hat{V}_{int}=\frac{pd}{r^3}\sum\limits_{i=1}^{{{N}_{m}}}{\left\{ {{{\hat{b}}}^{+}}\left[ {{{\hat{\sigma }}}_{i}}-3\left( {{{\hat{\sigma }}}_{i}}{{\cos }^{2}}{{\theta }_{i}}+{{{\hat{\sigma }}}_{\bot i}}\cos {{\theta }_{i}}\sin {{\theta }_{i}} \right) \right]+h.c. \right\}}\lb{sum_expl}
\eeq
Expression~\rf{sum_expl} can be simplified for a large number of absorbers $N_m\gg 1$. Summing over $i$, we approximate ${{\cos }^{2}}{{\theta }_{i}}\approx 1/2$ and neglect the product ${{{\hat{\sigma }}}_{\bot i}}\cos {{\theta }_{i}}\sin {{\theta }_{i}}$ repeatedly changing the sign with $i$. After such a simplification of  Eq.~\rf{sum_expl},  the Hamiltonian \rf{Ham_0} written in the interaction picture \ct{Landau2013_QM} and RWA approximation \ct{SLW} becomes
\beq
H=-{{\hat{a}}^{+}}\left( \hbar {{\Omega }_{p}}\hat{b}+\hbar {{\Omega }_{m}}\sum\limits_{i=1}^{{{N}_{m}}}{{{{\hat{\sigma }}}_{i}}} \right)-\hbar {{\Omega }_{int}}{{\hat{b}}^{+}}\sum\limits_{i=1}^{{{N}_{m}}}{{{{\hat{\sigma }}}_{i}}}+h.c.+\hat{\Gamma }\lb{Ham_simpl_2}
\eeq
where ${{\Omega }_{p}}={p{{E}_{0}}}/{\hbar }$, ${{\Omega }_{m}}={d{{E}_{0}}}/{\hbar }$ and ${{\Omega }_{int}}={pd}/2\hbar {{r}^{3}}$. The Hamiltonian  \rf{Ham_simpl_2} corresponds to the approximation when the polarisation of the absorber transitions is parallel to the polarisation of the external field. 
\subsection{Equations of motion}
We write Heisenberg the equations of motion $i\hbar \dot{\hat{A}}=\left[ \hat{A},H \right]$ where $\hat{A}$ is an operator from the Hamiltonian \rf{Ham_simpl_2}. The equation for the plasmonic oscillation Bose operator $\hat{b}$ is 
\beq
\dot{\hat{b}}=i{{\Omega }_{p}}\hat{a}+{{\Omega }_{int}}\hat{p}-\Gamma \hat{b}+(\sqrt{2\Gamma }){{\hat{b}}_{in}} \lb{eq_m_b}
\eeq
where we introduce the polarisation operator of all absorbers
\beq
\hat{p}=i\sum\limits_{i=1}^{{{N}_{m}}}{{{{\hat{\sigma }}}_{i}}}.\lb{pol_all_m}
\eeq
The last two terms are added to Eq.~\rf{eq_m_b} according to the input-output theory \ct{PhysRevA.46.2766,PhysRevA.30.1386}, they describe the dissipation of the dipole momentum oscillations with the rate $\Gamma$; ${{\hat{b}}_{in}}$ is the Bose operator of the bath. The stationary solution of Eq.~\rf{eq_m_b} is
\beq
\hat{b}=({i{{\Omega }_{p}}}/{\Gamma })\hat{a}+({{{\Omega }_{int}}}/{\Gamma })\hat{p}+(\sqrt{{2}/{\Gamma }}){{\hat{b}}_{in}}\lb{st_sol_b_eq}
\eeq
Using well-known properties of Dirac notations \ct{Lifshitz1982a}, we arrive at the commutation relations
\beq
\left[ {{{\hat{\sigma }}}_{i}},\hat{\sigma }_{i}^{+} \right]={{\hat{N}}_{0i}}-{{\hat{N}}_{1i}}\lb{com_rel_sigma}
\eeq
where ${{\hat{N}}_{0i}}={{\left| 0 0\right\rangle }_{i}}{{\left\langle  00 \right|}_{i}}$, ${{\hat{N}}_{1i}}={{\left| 0 1\right\rangle }_{i}}{{\left\langle  01 \right|}_{i}}$ are ith absorber state population operators. Operators with different indices $i$ are commute. Using the commutation relations \rf{com_rel_sigma}, Hamiltonian \rf{Ham_simpl_2}, and  polarisation \rf{pol_all_m} we write the equation of motion 
\beq
\dot{\hat{p}}=-\left( {{\Omega }_{m}}\hat{a}+{{\Omega }_{int}}\hat{b} \right)\left( {{{\hat{N}}}_{0}}-{{{\hat{N}}}_{1}} \right)-\gamma \hat{p}+\hat{F_p}\lb{pol_eq_moth}
\eeq
where ${{\hat{N}}_{\alpha }}=\sum\limits_{i=1}^{N}{{{{\hat{N}}}_{\alpha i}}}$, $\alpha =0,1$ are operators of populations of states of all absorbers. The last two terms in Eq.~\rf{pol_eq_moth} describe the absorber  polarisation decay with the rate $\gamma$ and the  noise with the Langevin force $\hat{F_p}$ of the absorber polarisation baths. 

Taking the commutation relations $\left[ {{{\hat{N}}}_{0i}},{{{\hat{\sigma }}}_{i}} \right]={{\hat{\sigma }}_{i}}$, Hamiltonian \rf{Ham_simpl_2}, and summing over all absorbers we obtain the equation of motion
\beq
{{\dot{\hat{N}}}_{0}}=\left( {{\Omega }_{m}}{{{\hat{a}}}^{+}}+{{\Omega }_{int}}{{{\hat{b}}}^{+}} \right)\hat{p}+h.c.+{{\hat{N}}_{1}}/\tau +{{\hat{F}}_{0}}\lb{N0_eq_m}
\eeq
where the last two terms  are due to the decay of the absorber excited state populations with the decay time $\tau$ and the Langevin force ${{\hat{F}}_{0}}$ corresponding to the population bath. The population operator of the absorber upper states  is  $\hat{N}_1=N_m-\hat{N}_0$.  
\section{The stationary absorption}\label{sec3}
From here we use the semi-classical approach as, for example, in the laser theory \ct{SLW}. We neglect fluctuations, drop the bath operators, and replace the operators with c-numbers (removing the hats from the notations). Calculating the semi-classical stationary solution of Eqs.~\rf{st_sol_b_eq} and \rf{pol_eq_moth} we find
\beq
p=-\frac{\left( {{{\Omega }_{m}}}/{\gamma }+{i{{\Omega }_{int}}{{\Omega }_{p}}}/{\Gamma \gamma } \right)a\left( {{N}_{0}}-{{N}_{1}} \right)}{1+\left( {{N}_{0}}-{{N}_{1}} \right)/{{N}_{int}}},\lb{st_p_sol}
\eeq
where ${{N}_{int}}=\gamma \,\Gamma /\Omega _{int}^{2}$.

Eq.~\rf{st_p_sol} helps us to understand the physical processes in the absorber-particle interaction.   The stationary polarisation of two-level absorbers in the resonant field without the particle is $p=-(\Omega_m/\gamma)a(N_0-N_1)$. 
The particle near the absorbers interacts with the external field at the rate $\Omega_p$, accepts the polarisation, re-emits the field, and through the re-emitted field "transfers" the polarisation (and the energy)  to absorbers with the rate $\Omega_{int}$. Such polarisation and the energy transfer is called plasmon induced resonance energy transfer (PIRET)   \ct{10.1063/1.5050209}. PIRET is suppressed by the polarisation decay: in the particle at the rate $\Gamma$ and in the absorbers at the rate $\gamma$. Therefore, the PIRET particle-absorber polarisation transfer rate  is $\sim i\Omega_{int}\Omega_p/\Gamma\gamma$, as we see in Eq.~\rf{st_p_sol}. The multiplier $i$ in the last expression and in Eq.~\rf{st_p_sol} is due to the well-known $\pi/2$ phase shift in the  resonant interaction of oscillators \ct{Lifshitz1982_M}. 

The interaction between the absorber and the particle is a two-way process: the particle transfers the polarisation to the absorber and vice versa: the absorber transfers the polarisation to the particle through the near field, causing a reduction in the polarisation of the absorber.  Such a reduction is described by the factor $1+\left( {{N}_{0}}-{{N}_{1}} \right)/N_{int}$  in the denominator of Eq.~\rf{st_p_sol}. 

To understand the physics and to characterise the stationary absorption  we write the  energy conservation law  using Eq.~\rf{N0_eq_m}: the rate $N_1/\tau$ of the energy relaxation from the upper states of the absorbers is equal to the sum of the photon absorption rate from the external field $\sim\Omega_m$ and the energy exchange rates at the interaction of the particle with the absorbers through the plasmonic mode $\sim \Omega_{int}$  
\beq
\frac{{{N}_{1}}}{\tau }=\left(-{{\Omega }_{m}} +\frac{{{i{\Omega }_{int}\Omega }_{p}}}{\Gamma }\right){{a}^{*}}p+c.c. -\frac{2\Omega _{int}^{2}}{\Gamma }\left\langle {{{\hat{p}}}^{+}}\hat{p} \right\rangle \lb{en_cons_law}, 
\eeq
where $\left<...\right>$ means  quantum mechanical averaging.  Considering  Eq.~\rf{pol_all_m}, the relation
\beq
\hat{\sigma}_i^+\hat{\sigma}_i = \hat{N}_{1i}\lb{dm_rel}
\eeq
and assuming a large number of absorbers $N_m\gg 1$ we write
\beq
\left\langle {{{\hat{p}}}^{+}}\hat{p} \right\rangle ={{N}_{1}}+\sum\limits_{i\ne j}{\left\langle \hat{\sigma }_{i}^{+}{{{\hat{\sigma }}}_{j}} \right\rangle }\approx {{N}_{1}}+{{\left| p \right|}^{2}}\lb{N1_extr}
\eeq
The term $N_1$ in Eq.~\rf{N1_extr} describes the spontaneous emission, and the second term describes a {\em collective} emission of absorbers to the plasmonic mode. These terms are similar to the cavity mode spontaneous emission and the collective emission terms in the laser theory \ct{Andre:19,Protsenko_2022}.  From Eqs.~\rf{st_p_sol}, \rf{en_cons_law} and \rf{N1_extr} we find
\beq
\frac{{{N}_{1}}}{\tau }=\displaystyle\left( \frac{2\Omega _{m}^{2}}{\gamma }+\frac{2\Omega _{p}^{2}}{\Gamma {{N}_{int}}} \right)\frac{n({{N}_{0}}-{{N}_{1}})}{1+\left( {{N}_{0}}-{{N}_{1}} \right)/{{N}_{int}}} -
\frac{2\Omega _{int}^{2}}{\Gamma }\left( {{N}_{1}}+{{\left| p \right|}^{2}} \right) \lb{en_cons_law2}
\eeq
where  $n=|a|^2$ is the mean number of photons  in the volume $V$ of the external field.  We will later express $n/V$ in terms of the external field intensity.

We introduce the absorption coefficients: ${{g}_{m}}={2\Omega _{m}^{2}}/{\gamma }$ of an absorber, ${{g}_{p}}={2\Omega _{m}^{2}}/{\Gamma }$ of the particle and   the spontaneous emission rate of an absorber to the plasmonic mode $\tilde{\gamma}_{sp}=2\Omega_{int}^2/\Gamma$. Using Eq.~\rf{st_p_sol} we find
\beq
\frac{2\Omega _{int}^{2}}{\Gamma }{{\left| p \right|}^{2}}=n\frac{\left( {{g}_{m}}+{{g}_{p}}/{{N}_{int}} \right){{{\tilde{\gamma }}}_{sp}}}{2\gamma }{{\left( \frac{{{N}_{0}}-{{N}_{1}}}{1+\left( {{N}_{0}}-{{N}_{1}} \right)/{{N}_{int}}} \right)}^{2}} \lb{polariz_1}
\eeq
and represent Eq.~\rf{en_cons_law2} in a compact form
\beq
\frac{{{N}_{1}}}{\tau }=n{{\tilde{g}}_{m}}({{N}_{1}})\frac{{{N}_{0}}-{{N}_{1}}}{1+\left( {{N}_{0}}-{{N}_{1}} \right)/{{N}_{int}}}-{{\tilde{\gamma }}_{sp}}{{N}_{1}}\lb{comp_en_cons}
\eeq
where the nonlinear absorption coefficient
\beq
{{\tilde{g}}_{m}}({{N}_{1}})=\left( {{g}_{m}}+\frac{{{g}_{p}}}{{{N}_{int}}} \right)\left[ 1-\frac{{{{\tilde{\gamma }}}_{sp}}}{2\gamma }\frac{{{N}_{0}}-{{N}_{1}}}{1+\left( {{N}_{0}}-{{N}_{1}} \right)/{{N}_{int}}} \right].\lb{nl_abs_coef}
\eeq
Using Eqs.~\rf{polariz_1} -- \rf{nl_abs_coef} we identify processes that affect the absorber excited state population.  The first process is the resonant absorption of the external field with the absorption coefficient $g_m$.   The second process is the resonant absorption enhancement  due to the nanoparticle near field. An external field energy is absorbed by the particle with the absorption coefficient $g_p$ and a part of this energy is transferred to the absorbers by the near field with the transfer coefficient $1/N_{int}$. Thus the linear (i.e. independent of $N_1$) part of the absorption coefficient ${{\tilde{g}}_{m}}$ is $g_m+g_p/N_{int}$. The non-linear term $\sim \tilde{\gamma}_{sp}$ in Eq.~\rf{nl_abs_coef} is due to the collective emission of absorbers to plasmonic mode. The reason for the suppression factor $1+(N_0-N_1)/N_{int}$ in the denominator of Eq.~\rf{comp_en_cons} is due to the transfer of the polarisation and the energy from the absorbers to the particle. The upper state population decays at the rate $\tilde{\gamma}_{sp}$ due to the spontaneous emission to the plasmonic mode, which is described by the last term on the right in Eq.~\rf{comp_en_cons}.  
\section{Plasmon-enhanced absorption conditions and  bistability}\label{sec4}
Resonant absorption in the plasmonic mode is a non-linear process. In addition to the well-known resonant transition saturation, the nonlinearity is caused by
by the collective interaction of the absorbers and the particle through the plasmonic mode, leading to an energy exchange between the absorbers and the particle. We will see how such nonlinear processes and the spontaneous emission to the plasmonic mode lead to the bistability of the absorption and affect the conditions for the best excitation of the absorbers. Let us introduce some parameters.  
\subsection{Parameters}
We express
\[
{{g}_{p}}n=\frac{2\Omega _{p}^{2}n}{\Gamma }=\frac{2n}{\Gamma }{{\left( \frac{p{{E}_{0}}}{\hbar } \right)}^{2}}=\frac{2n}{\Gamma }{{\left( \frac{p}{\hbar } \right)}^{2}}\frac{1}{{{\eta }^{2}}}\frac{2\pi \hbar {{\omega }_{0}}}{V}=\frac{4\pi n\omega_0\alpha_p}{\eta^2V}.
\]
Here we use ${{{p}^{2}}}/{\hbar \Gamma }={{\alpha }_{p}}$, where ${{\alpha }_{p}}$ is the absolute value of the resonant polarisability of the metal nanoparticle \ct{PhysRevA.71.063812}. The external field intensity is $
I=({c\eta }/{4\pi })\left\langle {{{\hat{\vec{E}}}}^{+}}\hat{\vec{E}} \right\rangle ={c\hbar {{\omega }_{0}}n}/{V\eta }$.
Here and below $c$ is the speed of light in vacuum. If we take $I=\hbar {{\omega }_{0}}P$ then the external field intensity is $P={cn}/{V\eta }$  in the number of photons per the unit of area per second. So the number of photons in the external field per the unit of volume is ${n}/{V}={\eta P}/{c}$ and 
\beq
{{g}_{p}}n={{\alpha }_{p}}\frac{4\pi {{\omega }_{0}}}{\eta }P=\frac{8{{\pi }^{2}}{{\alpha }_{p}}}{\eta {{\lambda }_{0}}}P={{\sigma }_{p}}P,\lb{abs_coef_n2}
\eeq
where ${{\lambda }_{0}}=2\pi /{{\omega }_{0}}$ is the wavelength of the absorbing field in vacuum, $\sigma_p$ is the absorption cross section of the particle resonance. Similarly, we find
\beq
{{g}_{m}}n=\frac{8{{\pi }^{2}}{{\alpha }_{m}}}{\eta {{\lambda }_{0}}}P={{\sigma }_{m}}P,\lb{gm_n}
\eeq
where $\alpha_m$ is the absolute value of the polarisability and  $\sigma_m$ is the absorption cross section of the two-level resonant absorber.  It is convenient to express the polarisability of the resonant absorber
\beq
\alpha_m=3\gamma _{sp}/4\eta k_{0}^{3}\gamma, \lb{abs_cr_sec} 
\eeq
in terms of the natural linewidth $\gamma _{sp}$ of the resonant transition \ct{Scully}; $k_0=\omega_0/c$ is the absorbed field wave vector in vacuum. Using the parameters introduced above, we express
\[
\frac{1}{{{N}_{{int}}}}\equiv\frac{\Omega _{{int}}^{2}}{\gamma\Gamma} =\frac{{{p}^{2}}}{\hbar \Gamma }\frac{{{d}^{2}}}{\hbar \gamma }\frac{1}{4{{r}^{6}}}=\frac{{{\alpha }_{p}}{{\alpha }_{m}}}{4{{r}^{6}}}, \]\[
\tilde{\gamma }_{sp}=\frac{2\Omega _{int}^2}{\Gamma }=\frac{\alpha_p\alpha_m}{2r^6}\gamma =\frac{2\gamma }{N_{int}}. 
\]
\subsection{Conditions for the absorption enhancement}
From the energy conservation law \rf{comp_en_cons} and \rf{nl_abs_coef} we derive
\beq
{{N}_{1}}={{\sigma }_{eff}}({{N}_{1}})P\tau \left( {{N}_m}-2{{N}_{1}} \right),\lb{N_1_eq0}
\eeq
where ${{\sigma }_{eff}}({{N}_{1}})$ is a non-linear (i.e. dependent on $N_1$) absorption cross section 
\beq
\sigma_{eff}(N_1)=\frac{\xi_{\sigma}\sigma_m}{[1+(N_m-2N_1)(\alpha_p\alpha_m/4r^6)]^2}.\lb{eff_abs_cs0}
\eeq
In Eq.~\rf{eff_abs_cs0} we use $N_0+N_1=N_m$ and introduce the linear  absorption cross section enhancement factor
\beq
\xi_{\sigma}=\frac{ 1+{\alpha _{p}^{2}}/{4{{r}^{6}}}  }{ 1+\gamma \tau \left( {\alpha _{p}\alpha_m}/{2{{r}^{6}}} \right) }. \lb{l_crssec_en_f}
\eeq
The nominator of \rf{l_crssec_en_f} is greater than 1 because of the direct energy transfer from the nanoparticle to the absorbing molecules. Such transfer is described by the factor $\alpha_p^2/4r^6$. The denominator of Eq.~\rf{l_crssec_en_f} is greater than 1 because of the spontaneous emission to the LPR mode. Such spontaneous emission reduces the absorption cross section.  $\sigma_{eff}$  is reduced relative to $\xi_{\sigma}\sigma_m$  due to the collective spontaneous emission of  absorbers to the LPR mode and the energy transfer back from absorbers to the particle. These two processes make the denominator in Eq.~\rf{eff_abs_cs0} greater than 1. 

The conditions, necessary for the absorption enhancement by the LPR follow from Eqs~\rf{eff_abs_cs0}, \rf{l_crssec_en_f}. The absorption enhancement is present when $\sigma_{eff}(N_1) > \sigma_m$. Assume at the beginning, that the denominator in Eq.~\rf{eff_abs_cs0} is close to 1, so that the collective emission to the LPR mode and the energy transfer from the absorbers to the nanoparticle are relatively small. This is the case at the saturation of the absorbing transitions, when $N_1$ is close to $N_m/2$, or at a small number of $N_m$ of absorbers, when
\beq
N_m{\alpha _{p}\alpha_m}/{2{{r}^{6}}}\leq 1. \lb{sm_n_abs}
\eeq
If Eq.~\rf{sm_n_abs} holds, the absorption cross section is enhanced at $\xi_{\sigma}>1$ which is when 
\beq
\alpha_p/\alpha_m > 2\gamma\tau,\hspace{0.5cm}\lb{cond_enh}
\eeq
or, if we insert the expression \rf{abs_cr_sec} for $\alpha_m$ into \rf{cond_enh}, when $\alpha_p>\alpha_p^{min}=3\gamma_{sp}\tau/2\eta k_0^3$. The relation \rf{cond_enh} is a necessary condition for the plasmon-stimulated absorption enhancement (found in the semiclassical approach).  Note that $2\gamma\tau\geq 1$, so $\alpha_p>\alpha_m$ is required for absorption enhancement. So absorption enhancement is possible if the nanoparticle absorbs the field better than an absorber, as mentioned for example in  \ct{https://doi.org/10.1002/adom.201700191}. 

However, the absorption of the nanoparticles must not be too high. Otherwise, the absorption by the molecules will be suppressed due to their collective emission to the LPR mode and the energy transfer from the molecules to the particle. Thus, by  combining conditions \rf{sm_n_abs} and \rf{cond_enh} we obtain the interval of $\alpha_p$, where the  absorption enhancement by LPR is possible
\beq
        2\gamma\tau\alpha_m \equiv \alpha_p^{min}\leq \alpha_p\leq\alpha_p^{max} \equiv 2r^6/N_m\alpha_m .\lb{interval} 
\eeq
Suppose that the decay and the dephasing of the upper absorber state are due only to the spontaneous emission to the vacuum modes and the emission to the LPR mode. Then $\gamma=\gamma_{sp}$, $2\gamma\tau = 1$ and the condition \rf{cond_enh} is reduced to $\alpha_p>\alpha_m=3/2\eta k_0^3$. We take the LPR wavelength $\lambda_{LPR} = 408$~nm for the spherical silver  nanoparticle in water ($\eta=1.33$)  and find $\alpha_m=3\cdot 10^{-4}$~mkm$^{3} \gg \alpha_p\approx 3\cdot 10^{-5}$~mkm$^{3}$ for the nanoparticle of the radius $r_p=13$~nm found for the highest LPR quality factor in the Appendix. The absorption enhancement is not possible with these parameters. Effective plasmon-stimulated absorption is possible for much smaller  $\alpha_m\leq 3\cdot 10^{-6}$~mkm$^{3}$ with a fast dephasing of the absorber transition (e.q. by some non-radiative processes), such that  $\gamma_{sp}/\gamma \leq 10^{-2}$. It is also necessary to provide $2\gamma\tau \ll 1$ for the absorption enhancement. Thus, for a typical value of $\gamma_{sp}\sim 1$~GHz, the absorber population must decay on a time scale $1-10$~ps  for effective LPR absorption enhancement. Physically, the absorber non-radiative population decay must be faster than the population decay due to the spontaneous emission to the LPR mode. Then the spontaneous emission to the LPR mode is only a small contribution to the total population decay and does not prevent the absorption enhancement.

It is interesting to note that {\em the fact} of the absorption enhancement does not depend on the distance $r$ between the absorbers and the particle (as long as the  absorbers are in the nanoparticle near the field zone). However, {\em the magnitude} of the enhancement does depend on $r$, as we will see.  
\subsection{ Bistability in the stationary absorption}
Suppose the number $N_m$ of absorbing molecules is such that $N_m{\alpha _{p}\alpha_m}/{2{{r}^{6}}}>1$. Then the absorption is strongly non-linear due to the collective effects, the energy transfer between the absorbers and the particle, which has a significant effect of the  $N_1$-dependent denominator in Eq.~\rf{eff_abs_cs0}.  The denominator decreases with $N_1$, so the absorption increases with $N_1$ for certain parameter values. As a result, the low and the high absorption regimes are possible,  overlapping at certain parameters, creating the {\em bistability} in the absorption. 

To study the bistability we introduce the dimensionless nanoparticle-absorber interaction  $A= {\alpha _{p}\alpha_m}{N}_{m}/{4{{r}^{6}}}$,  the external field intensity ${{P}_{eff}}=\xi_{\sigma}\sigma_mP\tau$ parameters, and consider 
the upper state population per absorber $n_1={N_1}/{N_m}$. With these parameters we write a cubic equation for $n_1$ 
\beq
{{{n}_{1}}}/({1-2{{n}_{1}}})+2{{n}_{1}}A+{{n}_{1}}\left( 1-2{{n}_{1}} \right){{A}^{2}}={{P}_{eff}},\lb{bist_eq}
\eeq
instead of Eq.~\rf{N_1_eq0}. Following a standard analysis of the catastrophe theory \ct{Poston1996}, we find that Eq.~\rf{bist_eq} has three  solutions when the number $N_m$ of absorbers is large enough that
\beq
A>A_{cr}=8.   \lb{bist_cond}
\eeq
Note that the condition \rf{bist_cond} is opposite to the small collective emission and the small absorber-to-particle energy transfer condition \rf{sm_n_abs}. Thus the collective emission and the absorber to particle energy transfer must be sufficiently large for the bistability. 

\begin{figure}[thb]\bc
\centering
\includegraphics[width=9cm]{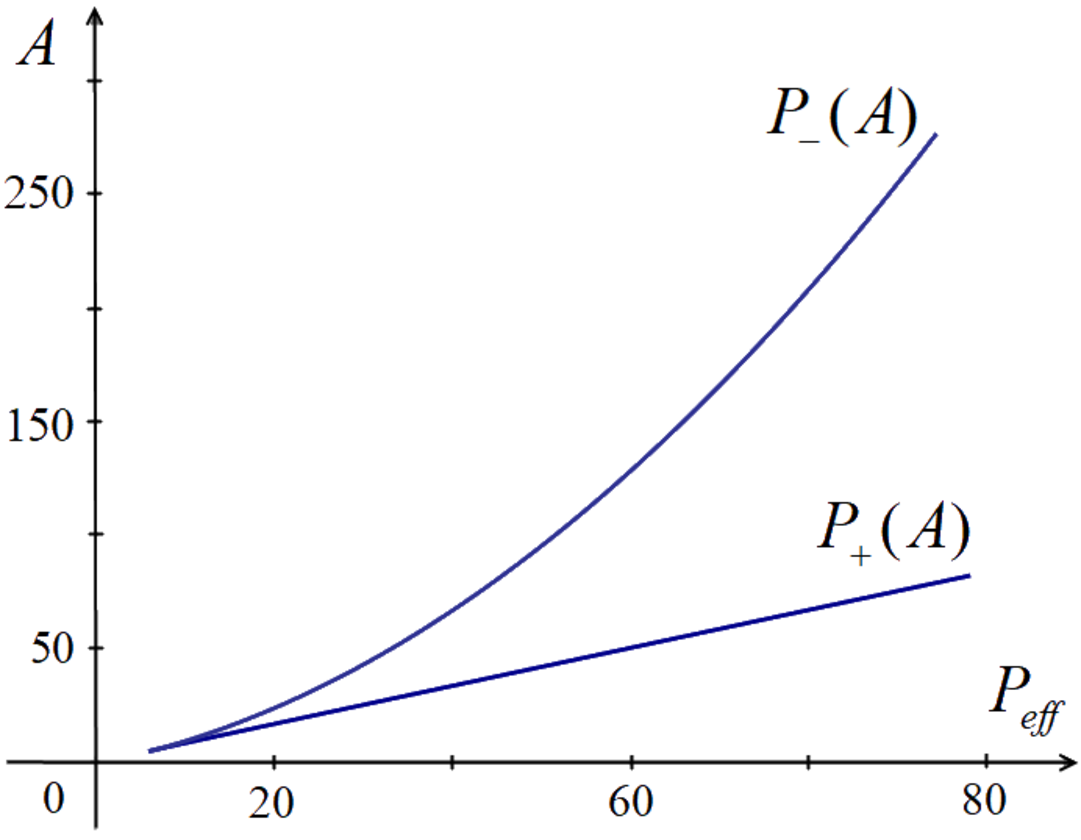}
\caption{Absorption bistability for values of $A$ and $P_{eff}$  lie between the curves $P_{\pm}(A)$.  }
\label{Fig2}\ec
\end{figure}

The  bistability domain is bounded by the $P_{\pm}(A)$ curves in  $A$, $P_{eff}$ parameter space 
\beq
P_{\pm}(A)={{{n}_{\pm }}}/({1-2{{n}_{\pm }}})+2{{n}_{\pm }}A+{{n}_{\pm }}( 1-2{{n}_{\pm }} ){{A}^{2}}\lb{bd_1}
\eeq
where ${{n}_{\pm }}(A)=\left( 3\pm \sqrt{1-8/A} \right)/8$. The bistability domain \rf{bd_1} is shown in Fig.~\ref{Fig2}. 
%
%
%
%
\section{Examples of absorption enhancement and the bistability}\label{sec5}
As an example, consider a spherical silver nanoparticle with a polymer shell in the water. The refractive index of most polymers is close to the refractive index of glass  $1.55$ \ct{Zhang:20}. The refractive index of water is $1.33$. For simplicity, we assume the same refractive index $\eta = 1.33$ for the nanoparticle shell and the water environment. 

The resonant polarisability of the nanoparticles is calculated in the Appendix, taking into account the radiative losses \ct{Bottcher1973,Protsenko_2012} and the losses due to collisions of the metal electrons with the nanoparticle surface \ct{KIK2007,Protsenko_2012}. We estimate the LPR wavelength $\lambda_{LPR} = 408$~nm and the half width $\Gamma\approx 80$~THz.  We take the nanoparticle radius $r_p=13$~nm, which corresponds to the maximum LPR quality factor $Q=29$.  The resonant absorbers are located on the nanoparticle dielectric shell of the thickness  $\Delta r \geq 5$~nm. We chose a minimum shell thickness, because of a rapid increase in absorber fluorescence quenching near metal nanoparticles has been observed at smaller $\Delta r$ \ct{PhysRevLett.96.113002,doi:10.1021/nl0480969}. Strong quenching at small $\Delta r$ can be caused by several mechanisms, besides the dipole-dipole interaction, such as  a nanometal surface energy transfer (NSET) \ct{GHOSH2015223}.  Taking $\Delta r \geq 5$~nm  we can neglect by the nanoparticle-absorber's interactions, but the dipole-dipole one, to keep our system as simple as possible.

The number of absorbers on the shell is large $N_m\geq 10$.  

We consider the stationary 
population  $n_1$ of the upper level of the absorber resonance transition as a property of the absorption. We compare $n_1$ with and without the nanoparticle, so we examine the LPR-stimulated absorption enhancement (or suppression) at different numbers of absorbers $N_m$, the dielectric shell thickness $\Delta r$, the absorber upper level depopulation time $\tau$ and the absorber transition half width $\gamma$.

\begin{figure}[thb]\bc
\centering
\includegraphics[width=7.5cm]{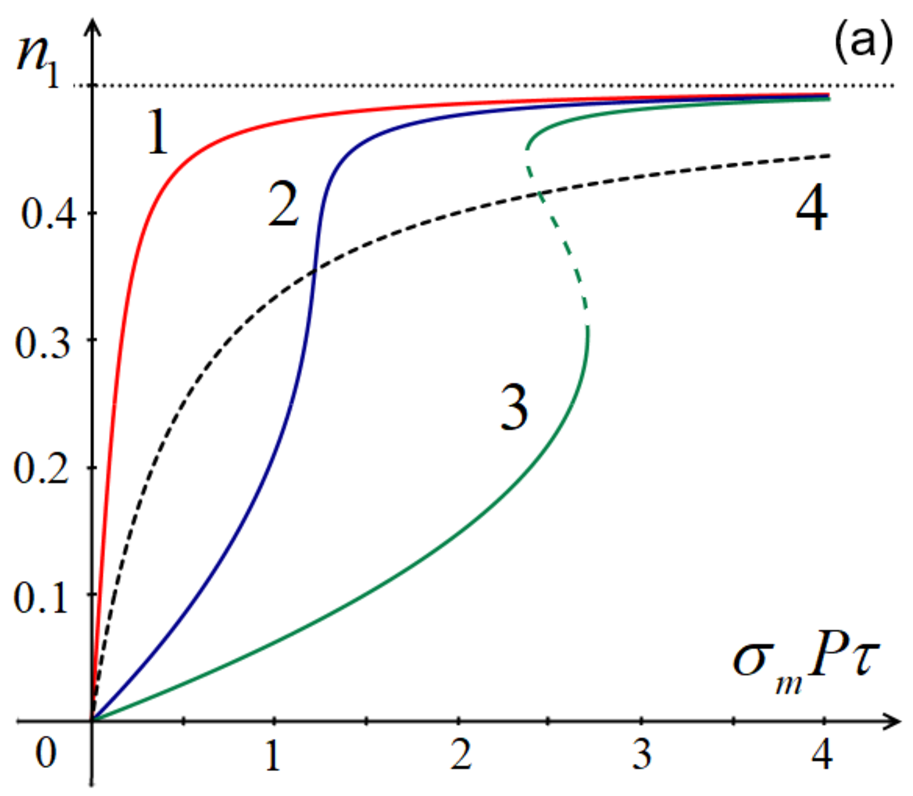}\hspace{0.5cm}\includegraphics[width=7.5cm]{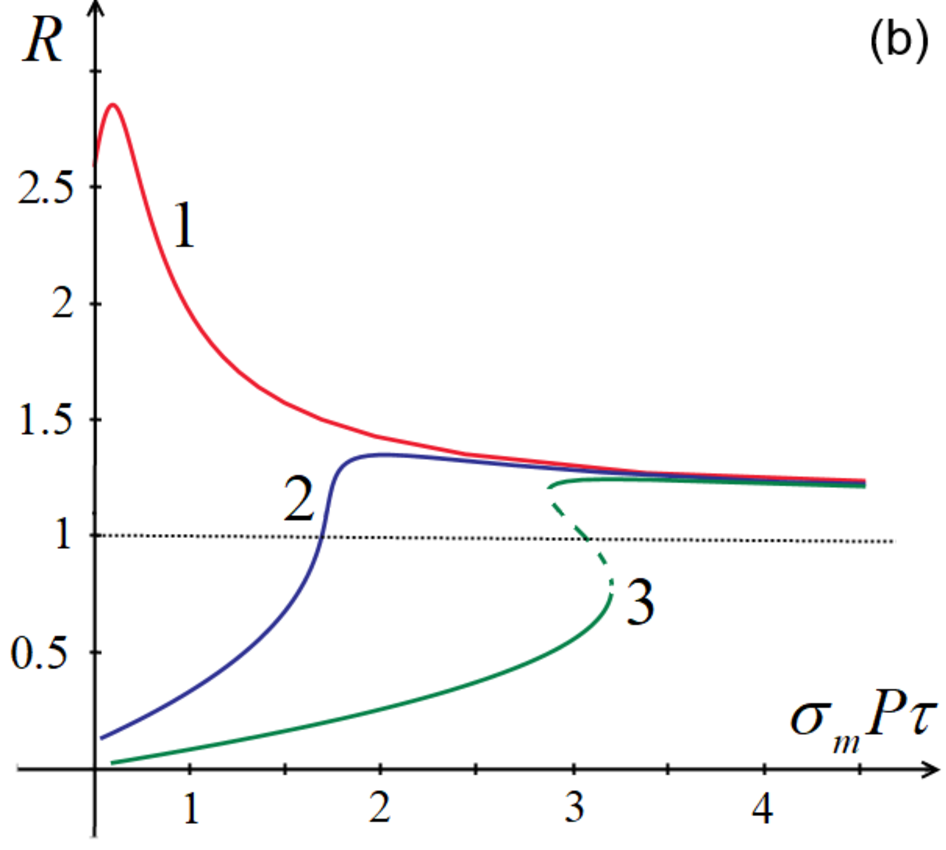}
\caption{(a) The  steady-state upper-level population per absorber versus the normalised external field intensity for different numbers of absorbers $N_m=10$ (curve 1), $80$ (2), and $120$ (3). Curve 4 is for the absorber upper-level population in free space. (b) The ratio $R$ of the absorber upper-level population with and without the nanoparticle (the LPR absorption enhancement) for the same parameters as in Figure~\ref{Fig3}a. The unstable steady-state absorption corresponds to the dashed parts of curves 3.}
\label{Fig3}\ec
\end{figure}

Suppose we have a fast depopulation of the upper absorber states with a population decay time $\tau=2$~ps. Such a depopulation determines the absorber transition half-width so $\gamma = 1/2\tau$. We take the parameter values written above and find that the condition \rf{cond_enh} for the LPR absorption enhancement is satisfied: $\alpha_p/\alpha_m=100>2\gamma\tau=1$. Figure~\ref{Fig3}a shows $n_1$, Figure~\ref{Fig3}b - the ratio $R=n_1/n_1^{(0)}$ at different $N_m$, as a function of the normalised external field intensity $\sigma_mP\tau$;  $n_1^{(0)}$ is an absorber upper level population calculated  without the nanoparticle.  At current parameter values, $\sigma_mP\tau=1$ corresponds to an external field intensity 0.54~MWt/cm$^2$. Such intensities can be achieved, for example, in a pulsed laser regime with pulse duration  $\sim 1$~ns $\gg \tau=2$~ps, so that  we have a quasi-stationary population of absorber states. The external field energy is about 0.5 mJ/pulse at a pulse cross section of 1~cm$^2$.   
%
%
%
%

Curves 3 in Figs.~\ref{Fig3}a,~b are for $N_m=10$, when $\alpha_p/\alpha_p^{max} \approx 1.8\sim 1$ and both conditions \rf{sm_n_abs} and \rf{cond_enh} (or \rf{interval}) are satisfied. We can see from curves 1 in Figs.~\ref{Fig3} that the LPR absorption is enhanced, relative to the free space absorption, at any input field intensity. The maximum enhancement is about 2.5 --3 times that of free space at a small field intensity far from the absorber transition saturation. 

We increase the number of absorbers.  Curves 2 in Figs.~\ref{Fig3} are for $N_m=80$ so  $\alpha_p/\alpha_p^{max} = 14.4\gg 1$ and the condition \rf{sm_n_abs} for a small collective emission and energy transfer from absorber to particle is not satisfied. We see the absorption {\em suppression} by LPR at small $\sigma_mP\tau <1.6$. Otherwise, there is the absorption enhancement for a large $\sigma_mP\tau$, when the collective spontaneous emission and the energy transfer are reduced due to the saturation of the absorber transition. Note that $A=0.97<A_{cr}$ for curves 2, so there is no absorption bistability. 

We increase $N_m$ further and take $N_m=120$ for curves 3 in Figs.~\ref{Fig3}, so $A=11.6>A_{cr}=8$, the condition \rf{bist_cond} is satisfied and we see the bistability in curves 3. The bistability is between regimes with the low and the high absorption regimes. The LPR suppresses the absorption at low external field intensities and enhances it at high external field intensities. The unstable steady state solution corresponds to the dashed parts of curves 3. 

%
%
\begin{figure}[thb]\bc
\centering
\includegraphics[width=7.5cm]{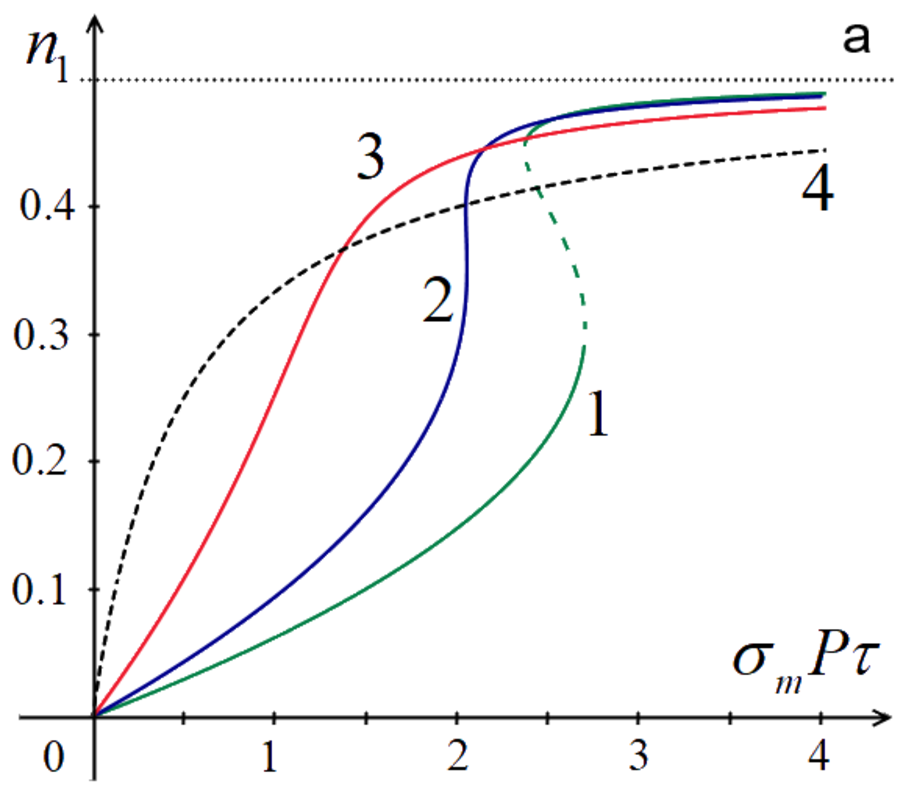}\hspace{0.5cm}\includegraphics[width=7.5cm]{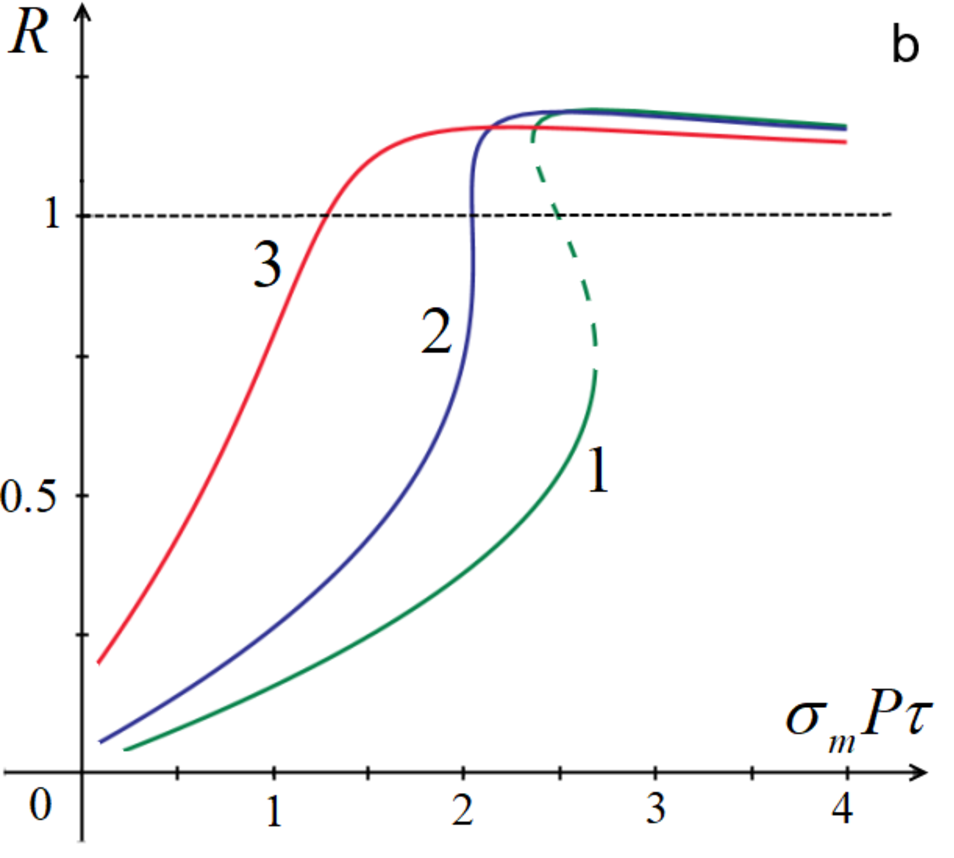}
\caption{ $n_1$ (a) and $R$ (b) versus the normalised input field intensity for $N_m=120$ and different  nanoparticle shell thickness $\Delta r=5$~nm (curves 1), $6$~nm (2) and 9~nm (3).  The rest of the parameters are the same as in Figure~\ref{Fig3}. The absorber without the nanoparticle corresponds to curve 4. The dashed parts of the curves represent the unstable steady-state solution. The bistability disappears with the increase of the shell thickness $\Delta r$.}
\label{Fig4}\ec
\end{figure}
%
%
Figure \ref{Fig4} shows the dependence of $n_1$ and $R$  on the normalised external field intensity $\sigma_mP\tau$ for a large number of absorbers $N_m=120$ and different  nanoparticle shell thicknesses $\Delta r = 5$ (curves 1), $6$ (2) and $9$ nm (3).  We see that the bistability disappears with only 1~nm increase in the nanoparticle shell thickness. For each shell thickness, the absorption is suppressed at low input field intensities and enhanced at high intensities, when the collective emission and the absorber-to-particle energy transfer are suppressed due to the absorber transition saturation. 

Let us see what happens with the decrease of the depopulation rate of the absorber state (i.e. the increase of  $\tau$).
%
%
\begin{figure}[thb]\bc
\centering
\includegraphics[width=7.5cm]{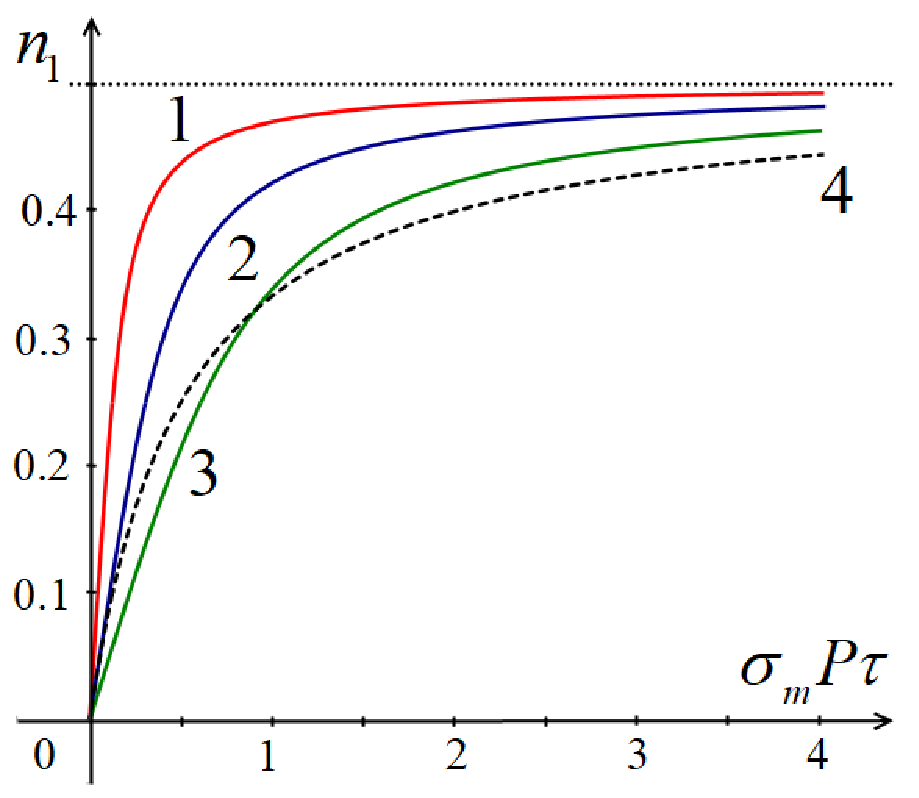}\hspace{0.5cm}\includegraphics[width=7.5cm]{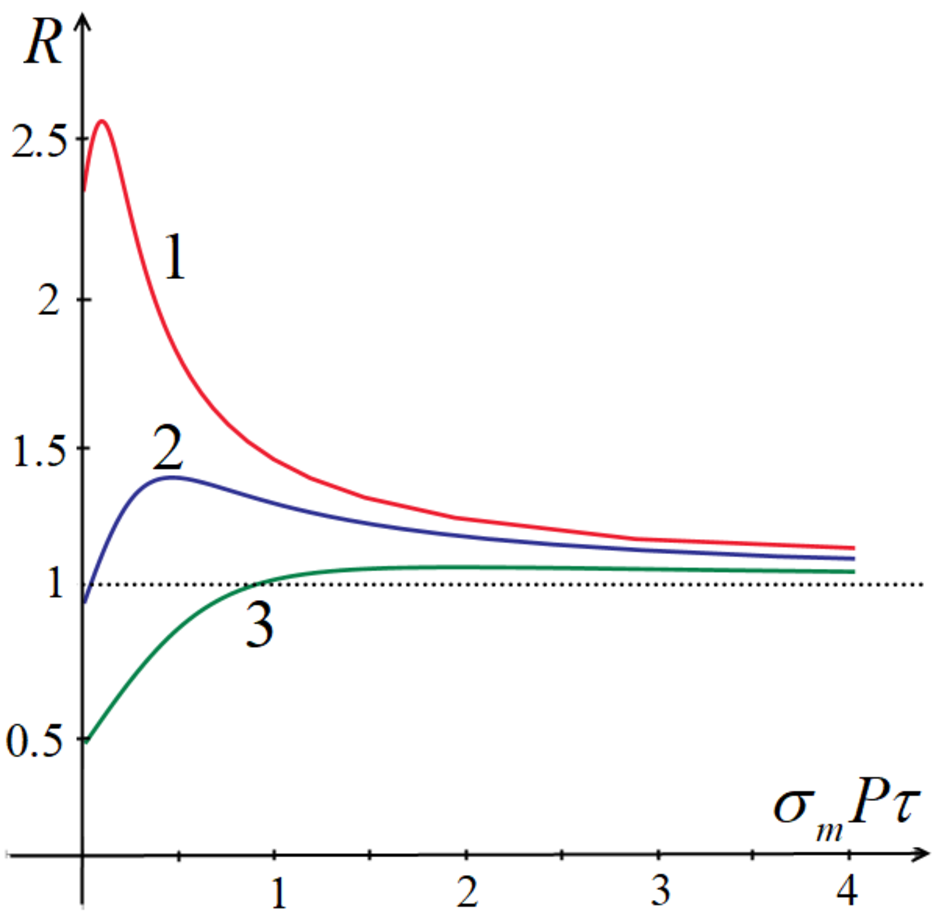}
\caption{ $n_1$ (a) and $R$ (b) versus the normalised input field intensity for $N_m=10$ and different absorber upper level depopulation time  $\tau =2$~ps (curves 1), $20$~ps (2) and 50~ps (3). Curve 4  is for an absorber without the nanoparticle. The shell thickness $\Delta r = 5$~nm, the rest of the parameters are the same as in Figure~\ref{Fig3}. The upper level population and the absorption are reduced with the increase of $\tau$. }
\label{Fig5}\ec
\end{figure}
%
%
We first  consider the case without the bistability and with a large absorption enhancement, as shown in Fig.~\ref{Fig3} by the curve 1 for $N_m=10$ and $\tau=2$~ps. The absorption enhancement by LPR is decreases with increasing $\tau$, from curves 1 to curves 3 in Fig~\ref{Fig5}, due to the increase of the spontaneous emission to the LPR contribution to the population decay. 

\begin{figure}[thb]\bc
\centering
\includegraphics[width=8cm]{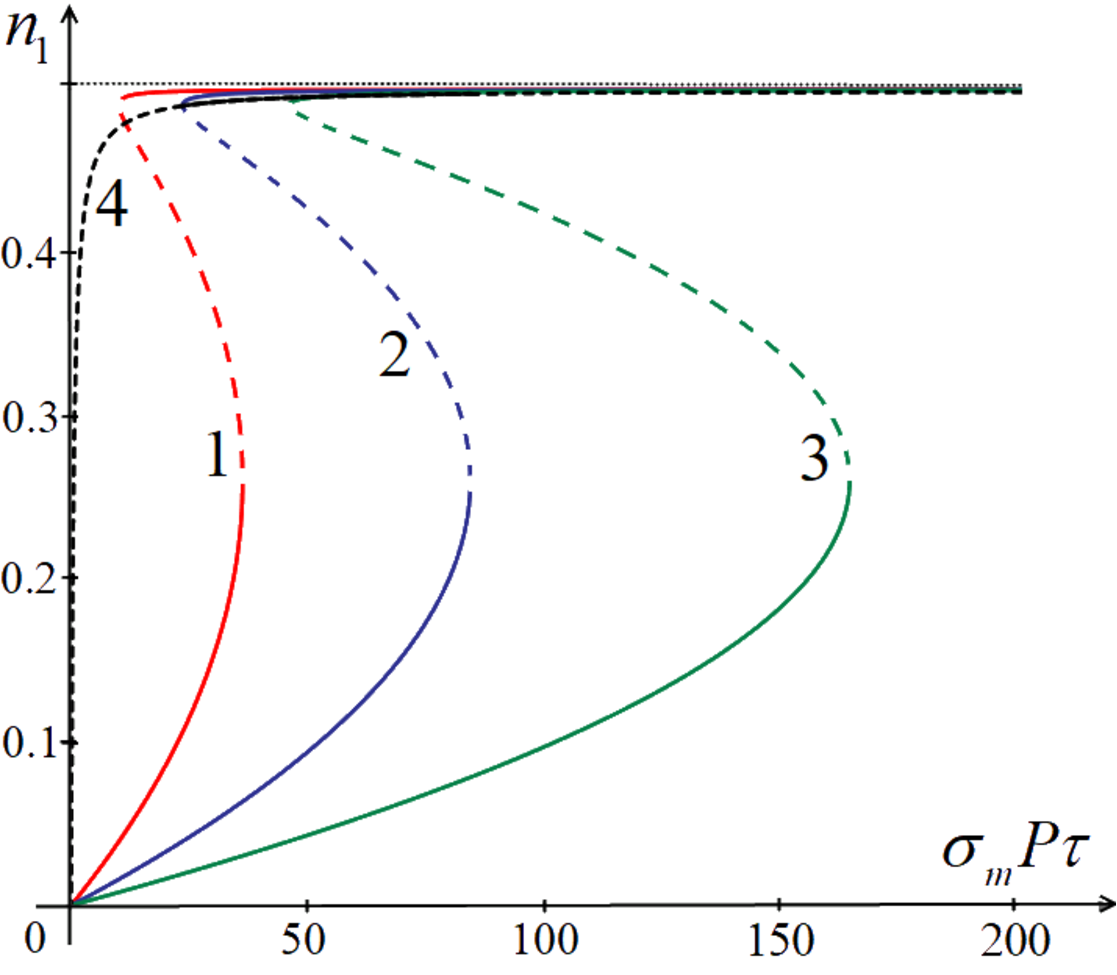}
\caption{The upper absorber state population $n_1(\sigma_mP\tau)$ at the absorption bistability, when $N_m=120$ and different $\tau=2$~ps (curve 1), $20$~ps (curve 2) and $50$~ps (curve 3). Unstable parts of the curves are indicated by the dashed lines. The bistability region increases as $\tau$ increases. Curve 4 is for the absorber in free space. The absorption enhancement (i.e. the positive difference in the upper parts of curves 1-3 and 4) is quite small.}
\label{Fig6}\ec
\end{figure}

Let us now increase the absorber state depopulation time $\tau$ at a large number of absorbers $N_m =120$, when the condition \rf{bist_cond} is satisfied and we have the absorption bistability. We see in Fig.~\ref{Fig6} that the bistability region grows with the increase of $\tau$, while the absorption enhancement is very small.
%
%
%

\begin{figure}[thb]\bc
\centering
\includegraphics[width=8cm]{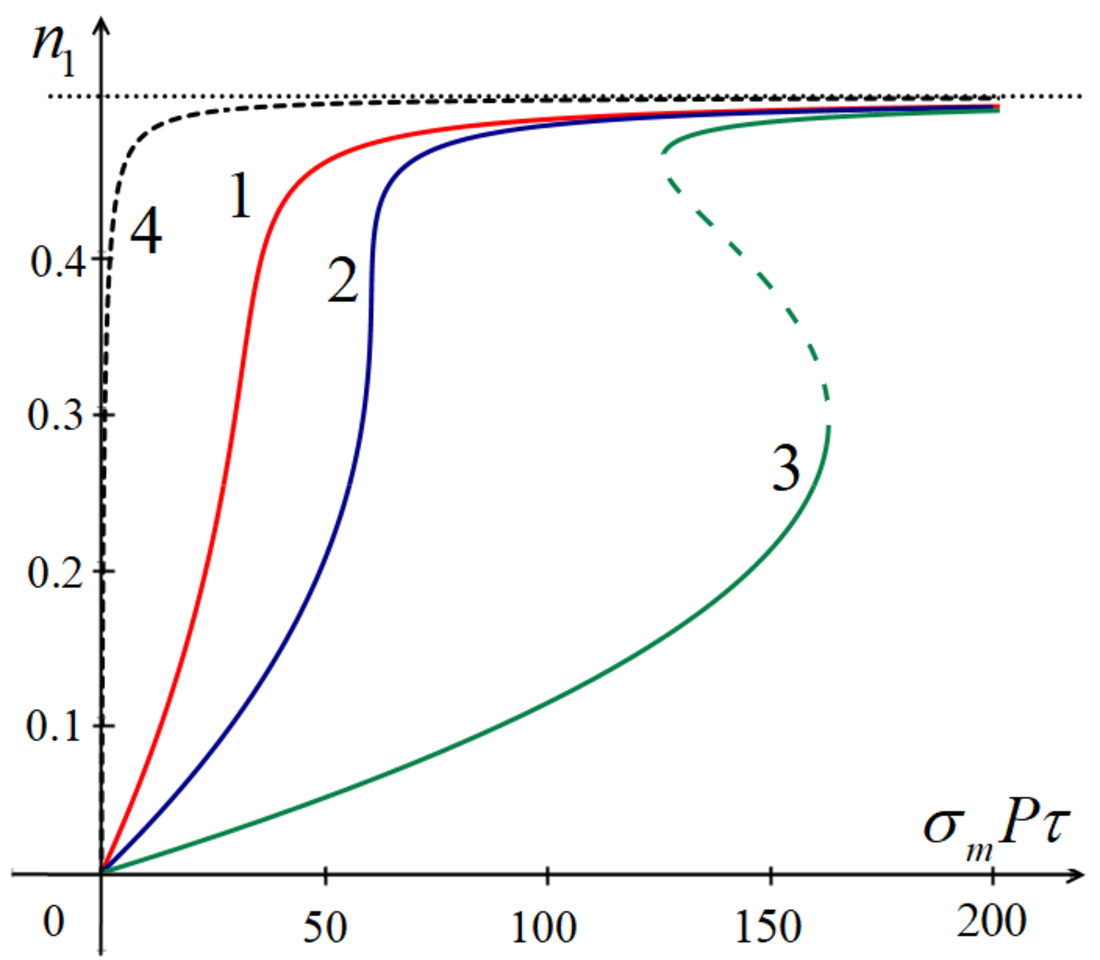}
\caption{The upper absorber state population $n_1(\sigma_mP\tau)$ with suppression of absorption by the spontaneous emission in the  LPR mode, when $\tau = 0.5$~ns and $\gamma = 100$~GHz. The number of absorbers $N_m=10$ is for curve 1; $N_m=16$ is for curve 2  and $N_m=30$ is for curve 3; curve 4 is for the absorber in free space. The bistability condition \rf{bist_cond} is satisfied and there is bistability in absorption exists for curve 3. The dashed part of curve 3 corresponds to the unstable steady state. There is no the absorption enhancement.}
\label{Fig7}\ec
\end{figure}

Suppose we have a large absorber upper state depopulation time $\tau=1/2\gamma_{sp}=0.5$~ns; such  depopulation occurs, for example, due to spontaneous emission into free space with the typical optical domain rate $\gamma_{sp}= 1$~GHz \ct{Scully}. We assume that the absorber transition has a half-width $\gamma = 100$~GHz. Then $\alpha_p/\alpha_m = 20 < 2\gamma\tau = 100$ so the condition \rf{cond_enh} is not satisfied and there is no absorption enhancement by the LPR at any external field intensity. Meanwhile, there is the bistability in the absorption, as shown by curve 3 in Figure~\ref{Fig7}, when 
the bistability condition \rf{bist_cond} is satisfied.
%
%
%
%
\section{Discussion}\label{sec6}
An   electric field enhancement
 near plasmonic nanoparticles increases the resonant absorption of atoms, molecules, semiconductor media, etc. The estimation \rf{l_crssec_en_f} shows the increase of the absorption cross section near plasmonic particle by a factor $\alpha_p^2/4r^6 = 9.5$ -- without taking into account the spontaneous emission of the absorber to the LPR mode (described by the denominator in Eq.~\rf{l_crssec_en_f}). Spontaneous and collective emissions to the LPR mode significantly reduce the absorption making it smaller than in the free space under some conditions.  In our examples, the maximum absorption enhancement near plasmonic nanoparticles is about a factor of 3 (see curve 1 in Figure~{\ref{Fig3}b}), taking into account the spontaneous emission to the LPR mode. 
 
 The maximum absorption enhancement is found for a small external field when the absorbing transition is not saturated. A noticeable absorption enhancement is possible with a fast (on ps time scale) depopulation of the upper absorbing state, when the rate of spontaneous emission to the LPR mode is close to or less than the depopulation rate due to other processes. When the population decay is not so fast, the relative suppression of the absorber excitation by the radiation to the LPR is large and we do not have the absorption enhancement, as shown in Figure~\ref{Fig7}. For LPR absorption enhancement, condition \rf{cond_enh} must be satisfied.
Spontaneous and collective emissions to the LPR mode are important factors that cannot be neglected in the resonant absorption near metal nanoparticles.

The enhancement of spontaneous emission by the cavity mode is well known in the laser theory \ct{Oraevskii:1994} as the Pursell effect \ct{PhysRev.69.674}. The metal nanoparticle is analogous to the cavity for the LPR mode, so such a cavity will enhance the spontaneous emission similar to a laser cavity. 

When a few resonant absorbers are placed close to metal nanoparticles,  collective emission from the absorbers into the plasmonic mode is possible. A photon emitted by the atom $i$ is re-absorbed by the atom $j$ and re-emitted  by the collective emission. This is a Dicke effect \ct{GROSS1982301}  with the probability (and hence the rate) is greater than for the photon emission from a single atom. The collective emission depopulates the upper state of the absorber.   Collective emission to the LPR mode is similar to the collective emission to a lasing mode \ct{Leymann,Jahnke,Andre:19, Protsenko_2022}. Collective emission is a non-linear process, its rate depends on the populations of the emitting states. In particular, collective
emission is suppressed at the saturation of the emitting transition.

The particle-absorber near-field interaction is a two-way process: the LPR mode of the metal nano-particle  excites the polarisation and transfers the energy through the near-field of the absorber,   and vice versa: absorbers excite the particle polarisation through the near-field. Such an energy transfer is the origin of the suppression factor in the denominator in Eq.~\rf{st_p_sol} for the absorber polarisation. The collective emission and the absorber-particle interaction depend on the absorber state population. 
Such a dependence leads to a positive feedback: the absorption is increased with the upper level population under  certain conditions. The feedback is fast: note the square in the non-linear denominator in Eq.~\rf{eff_abs_cs0}. Thus, for certain parameter values, the feedback  nonlinearity "overtakes" the transition saturation and leads to a {\em decrease} of the upper-level population $n_1$ with the external field intensity. As $n_1$ decreases, the steady-state  absorption is unstable, see the dashed parts of the curves in Figs.~\ref{Fig3}, \ref{Fig4}, and \ref{Fig6}. Such anomalous absorption causes absorption bistability. The bistability exists under the condition \rf{bist_cond}, which can be satisfied by the enhancement (when condition \rf{cond_enh} is true) or the suppression (when condition \rf{cond_enh} is not true) of absorption by LPR, see Figures~\ref{Fig4} and \ref{Fig6}.
In the case of LPR absorption enhancement, the bistability is between the absorption suppression and  absorption enhancement regimes, see Figures~\ref{Fig3} and \ref{Fig4}. 

Optical bistability is possible in plasmonic nanolasers with an external signal \ct{IEProtsenko_2008}. The model of \ct{IEProtsenko_2008} includes a single emitter near the plasmonic nanoparticle and does not consider the spontaneous and collective emissions to the LPR mode. 

The physical nature of the well-known absorptive optical bistability in a cavity with a two-level absorber (or absorbers) \ct{KOCH1985235} is the saturation of the resonant transition. This is different from the physics of the resonant absorption bistability in the LPR near metal nanoparticles described here. The absorptive bistability \ct{KOCH1985235} is not related to the collective emission and the near-field interaction of the absorbers with the metal nanoparticle. 

The nanoparticle with  the nearby resonant absorbers (molecules)   is probably one of the smallest systems with optical bistability, with a size much smaller than the optical wavelength.  

We have shown that the absorption enhancement near the metal nanoparticle and the bistability are strongly influenced by nonlinear collective absorption of the nanoparticle and nearby molecules, which occurs at high light power densities ($> 1~MW/cm^2$). Such power densities for single spherical metal nanoparticles can only be achieved with  powerful pulsed lasers. This makes  potential applications  in photovoltaics, photocatalysis, sensors, etc difficult. At the same time, the necessary power densities are easily achieved in plasmonic nanosystems with strong spatial localisation of the electromagnetic field in a plasmonic gap mode even at  moderate levels of external light excitation ($\sim 1~ W/cm^2$ and less) \ct{Baumberg2019,Yeshchenko2020,Huang2016,Lumdee2014}. This makes the study of the collective absorption near the plasmonic gap mode to be an interesting and important task. The present approach can be directly generalised to the plasmonic gap case.

Here we perform a semi-classical analysis of the resonance absorption near a metal nanoparticle. Only the term of the spontaneous emission in the LPR mode in Eq.~\rf{N1_extr} is introduced according to the quantum mechanical relation \rf{dm_rel}. We note that the term of the spontaneous emission into the laser mode appears in the same way in the laser theory  \ct{PhysRevA.50.4318,Andre:19}.  A more detailed theory of the resonant absorption near plasmonic nanoparticles must include quantum fluctuations of the field and the absorber state populations. Such a quantum analysis is necessary, taking into account a small volume of the plasmonic mode \ct{Maier2006},  holding only one or a few photons. Such an analysis will be performed in the future. 
\section{Conclusion}\label{sec7}

We study the enhancement of the resonant absorption of the electromagnetic field by two-level absorbers (atoms or molecules) in the vicinity of plasmonic nanoparticles, taking into account the spontaneous and collective emission from the absorbers to the localised plasmon resonance (LPR) mode of the particle. We derive quantum equations for the polarisation of the particle and the resonant absorber,  and  for the populations of absorbing states. We analyse the structure of the equations and describe the physical mechanisms of the stationary absorption and the particle-absorber interaction. 

We find that absorption enhancement by LPR is possible under certain conditions. Absorption suppression is also possible due to individual spontaneous  and collective emissions from absorbers to the LPR mode and energy transfer from absorbers to the particle. We find conditions necessary for absorption enhancement. One condition is a fast, on the picosecond time scale, depopulation of the excited absorber states, e.g. by some non-radiative processes. 

We find the conditions and study the bistability in the resonant absorption near the metal nanoparticle. The bistability is caused by the collective emission and the nonlinear interaction between the particle and the absorbers through the LPR near field. The bistability occurs with a strong external field excitation, a large number of absorbers in the vicinity of the nanoparticle, and is possible with absorption enhancement or suppression. 

The results can be applied wherever resonant absorption near metal nanoparticles is used, for example in plasmon-stimulated photocatalysis
\ct{Zhang_2013} or plasmonic photovoltaics \ct{Jang2016}.

\section*{Acknowledgments}
I.E.P., A.V.U. and N.V.N. acknowledge the Russian Science Foundation (Grant No.20-19-00559) for the support.
%
\section*{Appendix. Nanoparticle resonant polarisability}
The polarizability $\alpha_p$ of a spherical nanoparticle of the radius $r_p$ in an electric field of  wavelength $\lambda$, in a dielectric medium of refractive index $\eta$ is \ct{Protsenko_2012}
\beq
\alpha_p(\lambda,r_p) = r_p^3\frac{\varepsilon_p(\lambda,r_p)/\eta^2-1}{[1-iR_{rad}(\lambda,r_p)][\varepsilon_p(\lambda,r_p)/\eta^2+2]} \lb{nan_sp_pol}
\eeq
where $\varepsilon_p$ is the dielectric function of the nanoparticle material   and the factor 
\beq
R_{rad}(\lambda,r_p) = \frac{16\pi^3}{3}\left(\frac{\eta r_p}{\lambda}\right)^3\frac{\varepsilon_p(\lambda,r_p)/\eta^2-1}{\varepsilon_p(\lambda,r_p)/\eta^2+2} \lb{R_rad}
\eeq
describe the radiation losses.

The nanoparticle material dielectric function $\varepsilon_p(\lambda,r_p)$ taking into account the losses due to collisions of the  nanoparticle electrons with the nanoparticle boundary \ct{Protsenko_2012, Uskov2014} is
\beq
\varepsilon_p(\lambda,r_p)=\varepsilon_0-\frac{(\lambda/\lambda_p)^2}{(1+i\lambda/\lambda_c)(r_c/r_p+1)}.\lb{nanop_diel_f}
\eeq
The expression \rf{nanop_diel_f} with $r_c=0$ is a Drude dielectric function. For silver $\varepsilon_0=5$, $\lambda_p=0.14$~mkm, $\lambda_c=32$~mkm \ct{PhysRevB.91.235137}. The term $\sim r_c = A_0v_F\lambda_c/2\pi c$ describes the correction to the dielectric function due to collisions of electrons with the nanoparticle surface. The factor $A_0=0.7$, $v_F\approx 1.3\cdot 10^6$~m/s is the Fermi velocity of the electron, $c$ is the speed of light in vacuum \ct{Protsenko_2012}. Radiation losses increase with the nanoparticle radius $r_p$ while the losses due to collisions with the nanoparticle surface decrease with $r_p$. This, there is an optimum $r_p$, when the nanoparticle LPR has the maximum quality factor $Q_{LPR}$. Estimates show that the maximum $Q_{LPR} \approx 29$, when $r_p=r_p^{opt} = 12$~nm. The LPR wavelength $\lambda_{LPR} \approx 408$~nm and the LPR half-width $\Gamma \approx 80$~THz. The absolute value of the maximum resonant polarisability of the nanoparticles is  $\alpha_p= |\alpha_p(\lambda_{LPR},r_p^{opt})|\approx 3\cdot 10^{-5}$~mkm$^3$.

\bibliographystyle{elsarticle-num} 
\bibliography{myrefs}





\end{document}